\documentclass[aps,prb,twocolumn,showpacs,groupedaddress]{revtex4}
    \usepackage{amsmath, amsfonts,amssymb}
    \usepackage{graphicx}
    \usepackage{hyperref}
    \usepackage{color, soul}
\DeclareRobustCommand*{\citen}[1]{%
  \begingroup
    \romannumeral-`\x % remove space at the beginning of \setcitestyle
    \setcitestyle{numbers}%
    \cite{#1}%
  \endgroup
}
    
\begin{document}
\title{Kohn's localisation in disordered fermionic systems with and without interactions} 
\author{Vipin Kerala Varma and Sebastiano Pilati} \affiliation{The \textit{Abdus Salam} International Centre for Theoretical Physics, Trieste, Italy}
\date{\today}

\vspace*{-1cm}
\begin{abstract}
Understanding the metal-insulator transition in disordered many-fermion systems, both with and without interactions, is one of the most 
challenging and consequential problems in condensed matter physics.
In this paper we address this issue from the perspective of the modern theory of the insulating state (MTIS), which has already 
proven to be effective for band and Mott insulators in clean systems.
First we consider noninteracting systems with different types of aperiodic external potentials: uncorrelated disorder (one-dimensional Anderson model), 
deterministic disorder (Aubry-Andr\'e Hamiltonian and its modification including next-nearest neighbour hopping), and disorder with long-range correlations (self-affine potential).
We show how the many-body localisation tensor defined within the MTIS may be used as a powerful probe to 
discriminate the insulating and the metallic phases, and to locate the transition point.
Then we investigate the effect of weak repulsive 
interactions in the Aubry-Andr\'{e} Hamiltonian, a model which describes a recent cold-atoms experiment.
By treating the weak interactions within a mean-field approximation we obtain a linear shift of the transition point towards stronger 
disorder, providing evidence for delocalisation induced by interactions.
\end{abstract}


\maketitle
\section{Introduction}

In the modern theory of the insulating state (MTIS), which was initiated by the seminal article written by Kohn~\cite{Kohn}{}, the 
different behaviours of metals and insulators are attributed to the different organisations of the electrons in the many-body ground 
state.
 In insulators, the electrons satisfy a many-particle localisation condition~\cite{Kohnbook}{}. Kohn associated this localisation 
 with the break-up of the many-electron wave-function into terms which are 
localized in essentially disconnected regions of the many-particle configuration space.
%:
This approach is fundamentally different from conventional theories of insulators, which require the knowledge of, at least, 
low-lying excitations, and are tailored
towards a specific kind of insulator, depending on the physical mechanism which triggers the insulating behaviour. 
Some fundamental developments in the MTIS were achieved only in the late 1990's (thanks to the works by Resta and Sorella\cite{RS}{}, and 
others \cite{SWM,MV}). These developments began with the observation that the polarization is finite in insulator, while it is ill-defined 
in metals. This led to a definition of the localisation tensor \cite{RS} of a many-body system which is derived from the 
Berry-phase formulation of the polarisation as understood within the modern theory of the polarization~\cite{KingSmith, RV,Spaldin}.
 The direct connection between the many-body localisation tensor and the disconnected parts of the many-electron wave-function $-$ 
 as originally defined by Kohn $-$ was demonstrated by Souza, Wilkens and Martins \cite{SWM}.\\ 
The MTIS is supposed to be adequate to describe any kind of insulator, independent of the physical mechanism which induces 
the insulating behaviour. It should apply to band, Mott \cite{RS}, Anderson \cite{RestaAnderson}, quantum-Hall \cite{RestaTopological}
and possibly even to topological insulators. 
Therefore it represents a promising approach to address the outstanding open problem of the fate of Anderson localisation 
\cite{Anderson} in the presence of interactions~\cite{Gornyi,Fleishman, Basko}.
\\
So far, band and Mott insulators have been analysed within the framework of the MTIS \cite{RS}, both using lattice models in a 
tight-binding scheme, and also via \textit{ab initio} electronic structure simulations~\cite{Vetere,RestaAnderson,Veithen,Foulkes,Stella}.
Instead, Anderson insulators have received little attention. In particular, it is not known $-$ even for the noninteracting case $-$ 
whether the many-body localisation tensor may be used to
locate the critical point of the (Anderson) transition which separates the conducting and the insulating phases in disordered systems.\\
The first purpose of the paper is to investigate this issue in the noninteracting case. With this aim, we study the Anderson transition 
in noninteracting one-dimensional lattice models close to half filling.
Since disorder correlations play a fundamental role in low dimensional systems  (for instance, they determine the presence or 
absence of transition points and mobility edges~\cite{Izrlaev,dunlap1989absence,Pal}), we consider various models of disorder and 
study the many-body localisation tensor, and its reliability, in qualitatively different scenarios.
First, like Ref.~\citen{RestaAnderson}, we consider one-dimensional lattices with uncorrelated disorder, 
where the single-electron orbitals are expected to be exponentially localised at any nonzero disorder strength \cite{Anderson}. 
Then we focus on the more intriguing and instructive case of deterministic disorder 
due to an external periodic potential whose period is incommensurate with the lattice. All single-particle orbitals of 
this Hamiltonian (named Aubry-Andr\'{e} model) become localised, but only beyond a finite disorder strength~\cite{AubryAndre}. 
Next we consider a generalised Aubry-Andr\'{e} model including next-nearest neighbour hopping, where a 
mobility edge separating extended and localized single-particle orbitals was predicted~\cite{Biddle}.
Further, the case of non-deterministic disorder with tunable spatial correlations is addressed. In particular we consider a 
one-dimensional lattice with a self-affine disordered potential, where both localized and extended 
single-particle orbitals were suggested to be present \cite{Moura}.\\
The second purpose of this paper is to investigate the effect of interactions on the Anderson transition. In particular, we consider 
spin-1/2 fermions in the Aubry-Andr\'{e} model with on-site repulsive interactions. 
This model describes the experimental setup recently implemented with ultracold atomic gases in bichromatic optical 
lattices by the group of I. Bloch~\cite{Bloch}. We employ the Hartree approximation with temperature-annealed self-consistent iterations 
to determine the phase boundary separating the metallic and the insulating ground-states in the regime of weak interactions.\\
The paper is organized as follows: in Sec.~\ref{section1} we introduce the formalism of the MTIS, provide the definition of the 
many-body localisation tensor and describe our numerical procedure to compute it.
The analysis of the one-dimensional Anderson model is reported in Sec.~\ref{section2}. The Aubry-Andr\'e model is analysed in 
Sec.~\ref{section3}, and the generalised Aubry-Andr\'e model in Sec.~\ref{section4}.
The one-dimensional Anderson model with long-range correlated disorder is studied in Sec.~\ref{section5}.
In Sec.~\ref{section6} we address interaction effects in the Aubry-Andr\'e model.
In Sec.~\ref{section7} we draw the conclusions, focussing on the utility of the many-body localisation tensor 
to identify conductor-insulator transitions in disordered systems with and without interactions.

\section{Localisation tensor}
\label{section1}
 
In tune with Kohn's viewpoint on the origin of the insulating behaviour, the authors of Ref. \citen{RS} provided a quantitative 
definition of the many-body localisation tensor 
$\lambda_{\alpha \beta}$ (the indices $\alpha$ and $\beta$ indicate spatial directions) rooted in the modern theory of polarisation. 
This quantity measures the degree of localisation of the particles in the many-body ground state and permits to 
discriminate metallic and insulating phases.
In metals $\lambda_{\alpha \beta}$ is expected to be divergent in the thermodynamic limit, whereas it is finite in insulators; 
thus it defines a many-body localisation criterion for the ground state, referred to as Kohn's localisation~\cite{RestaAnderson}. 
The formula for $\lambda_{\alpha \beta}$ was originally obtained from the estimator of the polarization~\cite{RS}, and can also 
be derived using a general geometric quantum theory \cite{Resta}.
The cases of periodic and open boundary conditions need to be treated separately because the
 position operator is ill-defined in the former case \cite{RestaII}.\\
In the case of periodic boundary conditions the localisation tensor is obtained through the auxiliary quantity $z^{(\alpha)}_N$, 
which for a system of $N$ particles is defined as \cite{Resta, RS}

\begin{equation}
 \label{eq: zFactor}
 z^{(\alpha)}_N = \langle \Psi | \textrm{e}^{i\frac{2\pi}{L}\hat{\textbf{R}}_{\alpha}}|\Psi \rangle,
\end{equation}
where $\left|\Psi\right>$ is the many-body ground state, $\hat{\textbf{R}}_\alpha$ is the $\alpha$-component of 
the many-body position operator $\hat{\textbf{R}} = \sum_{i=1}^N\hat{r}_{i}$, where $\hat{r}_i$ is the position 
operator for particle $i$,  with the index $i=1,\dots,N$; $L$ is the linear system size.
We consider ground states of spin-$1/2$ fermions, with $N/2$ up and $N/2$ down spins.
For noninteracting particles (or mean-field schemes such as restricted Hartree-Fock \cite{RestaTopical}), $z^{(\alpha)}_N$ may be further simplified, 
giving 
\cite{RestaTopical, Resta} $z^{(\alpha)}_N = \textrm{det}^2\left[S^{(\alpha)}_{jj'}\right]$, where the matrix $\left[S^{(\alpha)}_{jj'}\right]$ (with the indices $j,j' = 1, 2, \dots N/2$) 
is the overlap matrix whose elements are given by
\begin{equation}
 \label{eq: Smatrix}
 S^{(\alpha)}_{jj'} = \int d\textbf{r}\phi^{*}_j(\textbf{r})\textrm{e}^{i\frac{2\pi}{L} r_\alpha}\phi_{j'}(\textbf{r}),
\end{equation}
where $\phi_j(\textbf{r})$ are the single-particle eigenstate spatial wave-functions ordered for increasing energies, and $\textbf{r}$ is the 
spatial coordinate. 
Using this auxiliary quantity, the localisation tensor is now defined as \cite{RS, RestaTopical}
\begin{equation}
 \label{eq: perLoc2}
 \lambda^2_{\alpha \beta} = -\frac{L^2}{4\pi^2N}\log{\frac{|z^{(\alpha)}_N||z^{(\beta)}_N|}{|z^{(\alpha \beta)}_N|}},
\end{equation}
where $z^{(\alpha \beta)}_N$ is defined as in Eq. \eqref{eq: zFactor} with $R_{\alpha}$ replaced by $R_{\alpha} - R_{\beta}$.
In one dimensional systems $N=L$ for half filling, and  the only component of the localisation tensor is the one corresponding to $\alpha = \beta = x$, which is 
  given by $\lambda^2_{xx} = -L\log{|z_N|}/2\pi^2$.

In the case of open boundaries, the position operator is well defined \cite{RestaII} and the localisation tensor may be evaluated 
according to the formula \cite{RS, Resta}
    \begin{equation}
     \label{eq: open}
     \lambda^2_{\alpha \beta} = (\langle \Psi|\hat{\mathbf{R}}_{\alpha}\hat{\mathbf{R}}_{\beta}|\Psi\rangle - \langle\Psi|\hat{\mathbf{R}}_{\alpha}|\Psi \rangle \langle\Psi|\hat{\mathbf{R}}_{\beta}|\Psi\rangle)/N.
    \end{equation}
For a system of independent 
electrons with $N/2$ spin-up and $N/2$ spin-down particles, this may be further simplified to give the squared localisation length as 
\cite{Resta}
\begin{equation}
 \label{eq: locOpen}
 \lambda^2_{\alpha \beta} = \frac{1}{N}\int d\textbf{r}d\textbf{r}'(\bf{r} - \textbf{r}')_{\alpha}(\textbf{r} - \textbf{r}')_{\beta}|P(\textbf{r},\textbf{r}')|^2,
\end{equation}
where $\rho(\textbf{r}, \textbf{r}') = 2P(\textbf{r},\textbf{r}')$ is the one-particle density matrix for a Slater determinant, which is given by \cite{Resta}
\begin{equation}
 \label{eq: openDM}
 \rho(\textbf{r}, \textbf{r}') = 2\sum_{j=1}^{N/2}\phi_j(\textbf{r})\phi^{*}_j(\textbf{r}').
\end{equation}
We stress that the length-scale $\lambda_{xx}$ is a
many-body localisation length. In particular, it is not simply related to the spatial extent of the single-particle eigenstates. 
For example, in the case of noninteracting band insulators,  $\lambda_{xx}$ is related to the spread of the maximally 
localised Wannier functions~\cite{SWM}, rather than to the Bloch wave functions. Notice that the latter (which are the single-particle eigenstates) 
are always extended.
There is no simple analogy with the Wannier functions for the case of disordered systems.\\
Further insight into the nature of $\lambda^2_{xx}$ can be obtained via substitution of Eq. \eqref{eq: openDM} into Eq. \eqref{eq: locOpen}. 
In the one-dimensional case, one obtains the expression~\cite{Veithen}:
\begin{eqnarray}
 \label{loc_expanded}
 \lambda^2_{xx} &=& \frac{2}{N}\sum_{i=1}^{N/2}  \left[ \langle \phi_i | \hat{x}^2 | \phi_i \rangle -  
 \langle \phi_i | \hat{x} | \phi_i \rangle^2 \right] \nonumber \\
 &-& \frac{2}{N} \sum_{i \neq j}   |\langle \phi_i | \hat{x} | \phi_j \rangle|^2 ,
\end{eqnarray}
where $\hat{x}$ is the single-particle position operator. The first sum in this equation is proportional to the second moment of the 
single-particle orbitals. 
The second sum in Eq. \eqref{loc_expanded} (which in the literature has been refereed to as the covariance term~\cite{Veithen}) 
originates from the antisymmetry of the many-particle wave-function, and would be absent in a single-particle analysis. It reflects the 
correlations between different orbitals. These two sums are of the same order of magnitude, as confirmed by inspection of numerical results. In particular, in the localized 
phase they both contribute to the value of $ \lambda^2_{xx}$, clearly indicating that this length scale reflects the properties of the 
many-particle wave-function, even in noninteracting systems.\\
In generic insulators, including those with correlations, $\lambda_{xx}^2$ is related to measurable quantities such as the 
mean-square fluctuations of the polarization and the inverse of the optical gap via a conductivity sum-rule \cite{SWM}. 
Furthermore, it is related to the spread of the generalized many-body Wannier functions, as defined in Ref.~\citen{SWM}, which are localized in disconnected regions of the high-dimensional configuration
space, establishing a direct connection with Kohn's theory of the insulating state.\\
The formulation to compute the localisation length proposed in Refs.~\citen{RS} and~\citen{Resta}, and briefly summarized in this section, 
provides a computational procedure to verify Kohn's 
contention that the many-body ground state contains sufficient information to ascertain whether the system is an insulator or a conductor, 
without recourse to the analysis of low-lying excitations.
 In this paper we provide evidence for a variety of disordered one-dimensional systems that this is indeed the case. 
 The saturation of $\lambda^2_{xx}$ in the thermodynamic limit is taken to signal Kohn's localisation \cite{RS, RestaAnderson}, 
 whereas its divergence indicates a conductor.\\
In our computations we consider both periodic and open boundary conditions and employ, respectively, equations \eqref{eq: perLoc2} and 
\eqref{eq: locOpen} to compute the localisation length. The single-particle spatial wave-functions $\phi_j(\textbf{r})$, needed to form the many-particle ground-state, 
are determined from full diagonalisation of the Hamiltonian matrix for a 
single spinless fermion using the Armadillo library \cite{Armadillo}{}.

  \section{1D Anderson model}
  \label{section2}
 We consider disordered tight-binding models of noninteracting spin-$1/2$ fermions defined by the Hamiltonian
\begin{eqnarray}
\label{eq: BasicHamiltonian}
  H = t\sum_{{r,\sigma}} (b_{{r,\sigma}}^{\dagger}b^{\phantom{\dagger}}_{{r}+{1},\sigma} + 
  \textrm{h.c})
  + W\sum_{r,\sigma} \epsilon_{{r}} {n}_{{r,\sigma}},
 \raggedleft
 \label{hamiltonian}
 \end{eqnarray}
\noindent
where ${r}=1,\dots,L$ is the discrete index which labels the lattice sites, $L$ is the linear system size,  
$b_{r,\sigma}$ ($b_{r,\sigma}^{\dagger}$) is the fermionic annihilation (creation) operator for a spin $\sigma=\uparrow,\downarrow$ particle at 
site $r$, and ${n_{r,\sigma}}=b_{r,\sigma}^{\dagger}b^{\phantom{\dagger}}_{r,\sigma}$ is the 
corresponding particle number operator. Here and in the rest of the article the lattice spacing is used as the unit of length, and the (even) total number of fermions $N$ is fixed.
The hopping amplitudes to the nearest neighbours are set by 
$t$, $\epsilon_{{r}}$ is the (random) value of the energy at lattice site ${r}$, while the parameter $W$ sets 
the strength of the disorder. \\

  \begin{figure}[ttp!]
\centering
\includegraphics[scale=0.735]{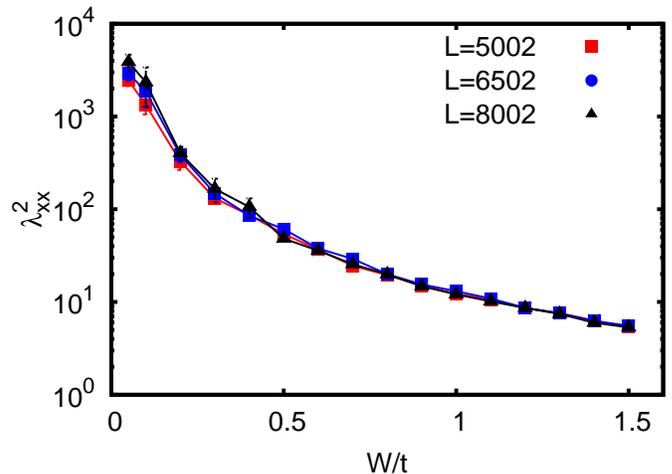}
\label{fig: 1DAndersonLocalisationLength}
     \caption{(Colour online) Half-filled 1D Anderson model with periodic boundary conditions: 
     Squared localisation length $\lambda_{xx}^2$ (log-scale) as a function of the disorder strength $W/t$. 
     Data for different systems sizes $L$ are shown, and their mutual agreement indicates Kohn's localisation at all disorder strengths. 
    The lines are guides to the eye. Here and in all figures the unit of length is the lattice spacing.}  
     \label{fig: 1DAnderson}
   \end{figure}
   
In this section we address, from a many-particle perspective, the Anderson model of localisation where the on-site 
energies $\left\{\epsilon_{{r}}\right\}$ are sampled from a uniform probability distribution in the interval $[-1, +1]$.\\
For noninteracting many-particle systems in the ground state, the wave function is the Slater determinant 
 formed with the lowest-energy occupied single-particle spin orbitals.
 The number of fermions per spin component determines the Fermi energy.
In this article, we consider the many-particle ground state of $N$ spin-1/2 fermions, with $N/2$ spin-up and $N/2$
spin-down particles.\\
We recall that in one dimensional (1D) systems with uncorrelated disorder all single-particle orbitals are localised - 
meaning that they are characterised by an exponentially decaying envelope -
for any nonzero disorder strength $W$. This follows from the one-parameter scaling theory \cite{Gang4}, and was 
also shown more rigorously in Refs.~\citen{Goldsheidt,Lifshits}.
According to Anderson's criterion of localisation~\cite{Anderson}, which is based on the localized shape of the single-particle orbitals at the Fermi energy, the system should be an insulator at any 
filling. \\
In Fig. \ref{fig: 1DAnderson}, we show the results for the squared localisation tensor $\lambda_{xx}^2$ as a function of the disorder strength $W/t$.
The data corresponding to three large (even) lattice sizes with periodic boundary conditions are shown. The lattices are half filled, and ensemble averaging of the results is 
performed considering $5 - 10$ realisations of the disorder pattern.
The localisation length $\lambda_{xx}$ varies by a few orders of magnitude as we tune the disorder strength.
However, it is always finite and system-size independent, in the whole range of disorder strengths we explore, which extends down to the
extremely weak disorder $W/t = 0.05$. These findings constitute a clear signature of Kohn's localisation. 
Also, the variation of $\lambda_{xx}^2$ with the disorder strength exhibits no sharp features (as opposed to the results of the next sections).
We verified that the data obtained using open boundary conditions (not shown) agree with those obtained using periodic boundary 
conditions.\\
Therefore, we conclude that the formalism of the MTIS predicts the many-particle ground-state of the 1D Anderson model to be an insulator, 
in agreement with the theory of Anderson localisation and the one-parameter scaling theory~\cite{Gang4}. 
However, in this latter formalism the insulating character is attributed to the localised 
shape of the single-particle orbitals in the vicinity of the Fermi energy, while the MTIS deals with the many-body ground-state 
wave-function.

%%%%%%%%%%%%%%%%%%%%%%%%%%%%%%%%%%%%%%%%%%%%%%%%%%%%%%%%
%%%%%%%%%%%%%%%%%%%%%%%%%%%%%%%%%%%%%%%%%%%%%%%%%%%%%%%%
\section{Aubry-Andr\'{e} model} 
  \label{section3}
   \begin{figure}[bbp!]
      \includegraphics[scale=1.75]{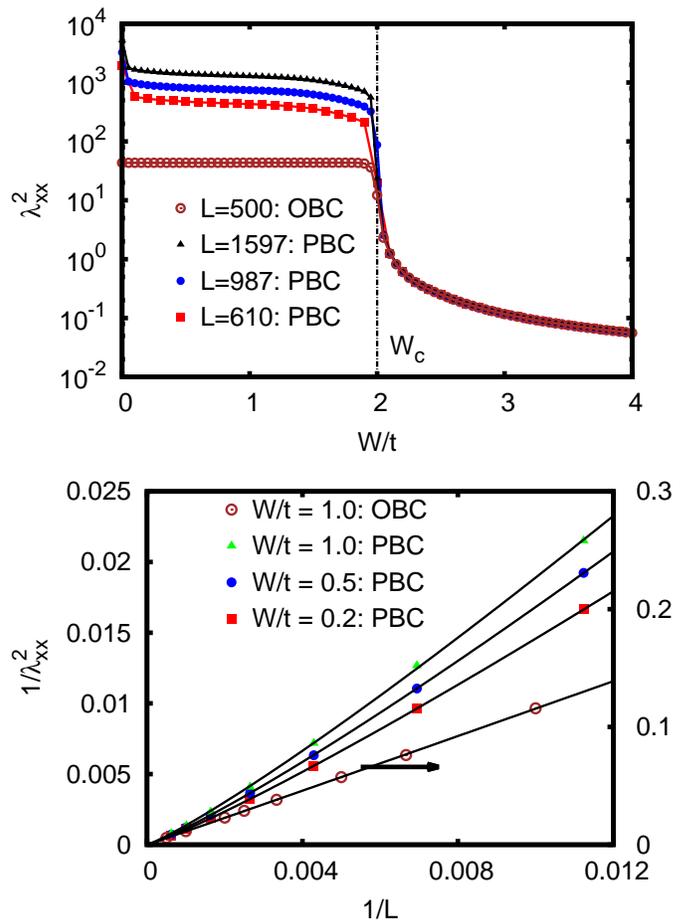}
     \caption{(Colour online) Half-filled 1D Aubry-Andr\'{e} model. 
    Top panel: squared localisation length $\lambda_{xx}^2$ (log-scale) as a function of the quasi-disorder strength $W/t$ for periodic (PBC) and open (OBC) boundary conditions, 
    and for different Fibonacci lengths $L$ of the chain; 
    the vertical line indicates the critical disorder strength $W_c/t = 2$, separating extended from localised states.
    Bottom panel: Scaling of $\lambda^{-2}_{xx}$ with the inverse system size $1/L$ in the conducting phase $W_c/t < 2$. 
      The continuous curves represent the power-law fits $\lambda^{-2}_{xx} = cL^{-\gamma}$ ($c$ and $\gamma$ are fitting parameters).
      $\lambda^{-2}_{xx}$ diverges with the exponent $\gamma \approx 1.14$ for PBC (left axis) and with $\gamma = 1.008(5)$ 
      for OBC (right axis).}
     \label{fig: NativeAubryAndre}    
   \end{figure}

In this section we consider the one-dimensional Aubry-Andr\'{e} model \cite{AubryAndre}. This is described by the 
Hamiltonian defined in Eq. \eqref{hamiltonian},  but with the on-site energies given by the incommensurate potential
$\epsilon_{{r}}=\cos(2\pi {{r}} g + \theta)$, where $g = (\sqrt{5} + 1)/2$ is the golden ratio, and $\theta$ is an (almost) arbitrary phase.
This is an archetypal model to study Anderson transitions in lower dimensions; it has been experimentally 
realized with ultracold atomic gases trapped in bichromatic optical lattices~\cite{Roati}, and also in quasi-periodic 
photonic lattices~\cite{Lahini}.\\
The sinusoidal potential does not display periodicity on a finite lattice, and so the Aubry-Andr\'{e} model is, in this sense, disordered. 
However, this disorder is deterministic, and so is not truly random.
In the presence of such ``deterministic disorder'', as opposed to 
true disorder, the one-parameter scaling theory of Ref. \citen{Gang4} does not apply.
In fact, it is known that this model hosts a transition from a diffusive phase at weak disorder, to a localised phase at strong disorder. 
In the former phase all single-particle eigenstates are extended over the whole system (possibly with the exception of a zero-measure set
 of non-exponentially localised states)~\cite{AubryAndre}. In the latter phase they are all localised if $g$ is a Diophantine number 
 (which is the case considered here) and for almost every value of $\theta$~\cite{jitomirs}. We chose $\theta = 0$. The transition occurs at the critical disorder strength $W_c/t = 2$.\\
In simulations with periodic boundary conditions we need to consider system sizes given by Fibonacci numbers, so that the potential fits the 
periodicity of the lattice. 
The results for the squared localisation length $\lambda^2_{\textrm{xx}}$ of half-filled lattices are 
displayed in Fig. \ref{fig: NativeAubryAndre} (top panel), both for periodic and open boundary conditions.
 A sharp variation of $\lambda^2_{\textrm{xx}}$ occurs in the close vicinity of $W/t = 2$.
For $W/t > 2$, the localisation length is finite and does not depend on the system size. This is a signature of Kohn's localisation.
Instead, for  $W/t < 2$, a very rapid increase of $\lambda^2_{\textrm{xx}}$ with the system size is observed, possibly indicating a metallic phase.
In order to confirm this supposition we perform a detailed analysis of the finite-size scaling of $\lambda^2_{xx}$. Various datasets 
obtained in the regime $W < 2$ are shown in Fig. \ref{fig: NativeAubryAndre} [bottom panel]. A best-fits analysis indicates that these data are accurately 
described by the (empirical) power-law fitting functions: $\lambda^{-2}_{xx} = cL^{-\gamma}$, where $c$ and $\gamma$ are the fitting parameters.  
The exponents obtained from the fitting procedure are $\gamma = 1.135(2), 1.151(2), 1.14(1)$ (for $W/t = 0.2, 0.5, 1$), and 
$\gamma = 1.008(5)$ ($W/t = 1$) for periodic and open boundary conditions, respectively. This fitting function predicts a divergence of the many-body localisation 
length in the thermodynamic limit, providing a clear indication of metallic behaviour. 
The divergence occurs both for periodic and open boundary conditions, but it is more rapid in the former case.\\
It is worth noticing that in the insulating phase the values of $\lambda^2_{\textrm{xx}}$ obtained using periodic and 
open boundary conditions are indistinguishable within our numerical accuracy. 
This independence from the type of boundary conditions is indeed expected for insulators, since in these systems the localisation 
lengths (and the polarisation) are bulk properties, as opposed to metals where they depend on the size of the 
system.\\
The analysis of the Aubry-Andr\'{e} model in the framework of the MTIS provides a clear signature of the metal-insulator transition at $W/t = 2$, in agreement with 
the Anderson criterion of localisation, which also predicts a phase transition at the same disorder strength since the single-particle orbitals change from extended to localised~\cite{AubryAndre}.
 We point out that we also performed a similar analysis of the Aubry-Andr\'{e} model at different lattice fillings in the regime $0.1
 < N/(2L)<0.9$, 
 without observing measurable shifts of the critical point. This is also expected following Anderson's criterion of localisation, 
 since the single-particle spectrum of this model does not host mobility edges~\cite{AubryAndre}.
 %

%%%%%%%%%%%%%%%%%%%%%%%%%%%%%%%%%%%%%%%%%%%%%%%%%%%
%%%%%%%%%%%%%%%%%%%%%%%%%%%%%%%%%%%%%%%%%%%%%%%%%%%
\section{Generalized Aubry-Andr\'{e} model} 
\label{section4}
    \begin{figure}[ttp!]
    \includegraphics[scale=1.75]{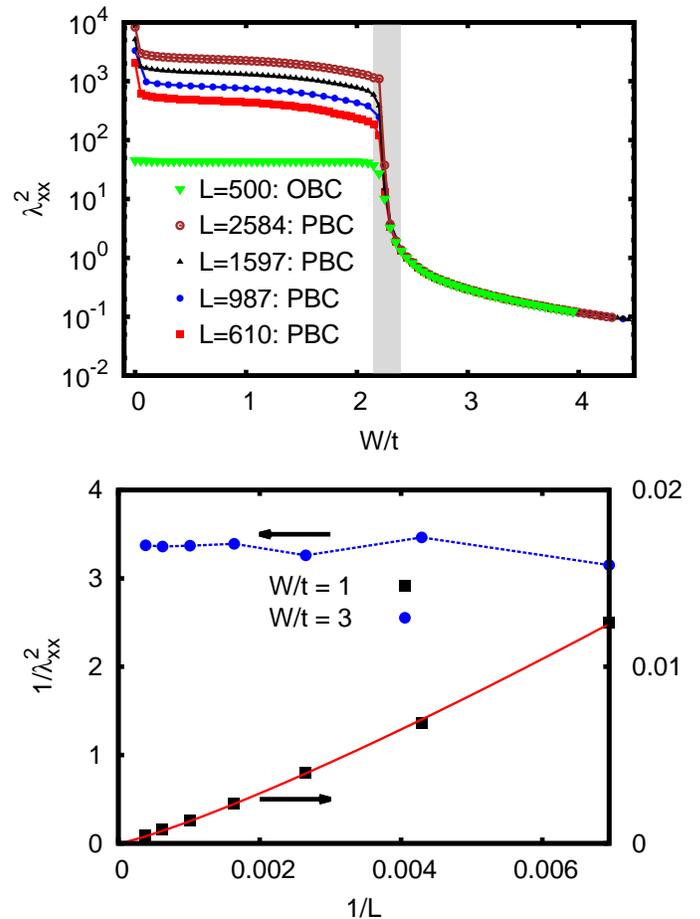}
    \caption{(Colour online) Generalized 1D Aubry-Andr\'{e} model with next-nearest hopping $t_2/t = 0.5$, at half filling. 
   Top panel: squared localisation length $\lambda^2_{xx}$ (log-scale) as a function of the quasi-disorder strength $W/t$. 
   Data for periodic (PBC) and open boundary conditions (OBC) are shown, for different chain lengths $L$. 
   The gray vertical stripe indicates the approximate location of the critical point between the metallic and the insulating phases.
   Bottom panel: scaling of inverse squared localisation length $1/ \lambda^2_{xx}$ with the inverse system size $1/L$ 
   at quasi-disorder strength in the metallic phase $W/t = 1$ and in the insulating phase $W/t = 3$ (blue dashed lines are a guide to eye).
   The continuous red curve represents the power-law fitting function $\lambda^{-2}_{xx} = cL^{-\gamma}$, 
   with the best-fit parameter $\gamma = 1.19(2)$.}
    \label{fig: ExtendedAubryAndre}    
  \end{figure}
  
  \begin{figure}[ttp!]
    \includegraphics[scale=0.7]{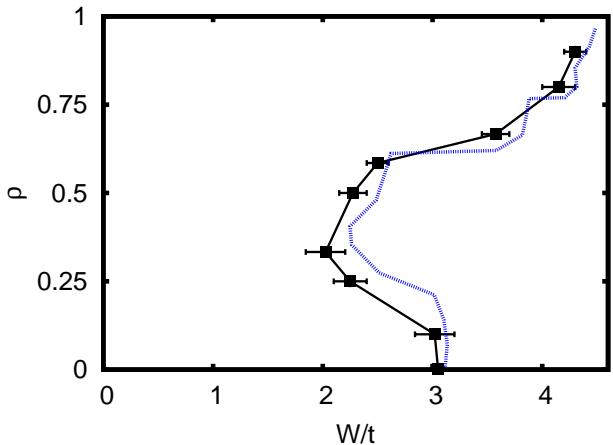}
    \caption{(Colour online) Zero-temperature phase diagram of the generalised Aubry-Andr\'{e} model with $t_2/t = 0.5$, 
    as a function of filling $\rho = N/(2L)$ and quasi-disorder strength $W/t$.
     The black points indicate the phase boundary separating the metallic phase (left) from the insulating phase (right) 
     obtained within the MTIS.
   Our results may be compared with the critical point extracted from the inverse participation data of Ref. \citen{Biddle}{} 
   (blue dashed curve).}
   \label{fig: ExtendedAAmobilityEdge}
  \end{figure}
Hopping processes beyond nearest-neighbour sites can dramatically alter the localisation properties, even causing the occurrence of single-particle mobility edges when none 
existed in the absence of such effects \cite{Biddle}{}.  
In this section, we consider the generalised Aubry-Andr\'{e} model, including next-nearest neighbour hopping.
With this modification, one obtains the Hamiltonian 
$H' = H + t_2 \sum_{{r},\sigma}  (b_{{r},\sigma}^{\dagger}b^{\phantom{\dagger}}_{{r}+{2},\sigma} + 
  \textrm{h.c})$, where $H$ is defined in Eq.~\eqref{hamiltonian}, $t_2$ is the energy associated to hopping to next-nearest neighbours, and the on-site energies $\epsilon_{\mathbf{r}}$ are 
  defined by the same incommensurate sinusoidal potential 
  of the native Aubry-Andr\'{e} model considered in the previous section.\\
This generalised Aubry-Andr\'{e} model
 was studied in Ref.~\citen{Biddle}. The analysis of the single-particle spectrum based on calculations of the inverse participation ratio (which measures the spatial 
 extent of the single-particle orbitals) presented evidence of the presence of mobility edges in a certain regime of disorder strength $W$.
  The location of the mobility edges was found to vary with $W$.
As in previous sections, here we analyse the many-particle ground-state of the generalised Aubry-Andr\'{e} Hamiltonian in the framework of the MTIS. 
We consider various lattice fillings, varying from vanishing density to full filling.
We fix the next-nearest neighbour hopping at  $t_2/t = 0.5$, a value which was also considered in Ref.~\citen{Biddle}.
An illustrative example of the dependence of the squared localisation length $\lambda_{xx}^2$ as a function of $W$ is shown in 
Fig. \ref{fig: ExtendedAubryAndre} (top panel), where the datasets correspond to half-filled lattices of different sizes $L$.
Here too, as in the case of the native Aubry-Andr\'{e} model, a sharp variation of $\lambda_{xx}$ occurs at a finite disorder strength $W_c$.
For $W > W_c$, $\lambda_{xx}^2$ is finite and system-size independent, indicating Kohn's localisation. 
Instead, for $W < W_c$, $\lambda_{xx}^2$ rapidly increases as $L$ increases. 
In order to assert whether in this regime the ground-state is metallic, we analyse the 
finite-size scaling of $\lambda_{xx}^2$ (see Fig. ~\ref{fig: ExtendedAubryAndre} [bottom panel]).
The scaling of the squared localisation length with the system size $L$ turns out to be accurately described by the empirical 
fitting function $\lambda^{-2}_{xx} = cL^{-\gamma}$, where $c$ and $\gamma$ are fitting parameters. At $W/t = 1$, the best fit is obtained with 
$\gamma = 1.19(2)$. 
This scaling behaviour clearly indicates a divergence of the localisation length, which is a signature of metallic behaviour.\\
In order to approximately pinpoint the phase boundary between the metal and the insulator, we determine the largest disorder strength 
where $\lambda_{xx}$ clearly 
diverges in the thermodynamic limit, and the smallest value of $W$ where it is system size independent, within numerical accuracy. 
This allows us to provide a (rather narrow) interval containing the critical disorder strength $W_c$. For the case of half-filling we 
obtain $W_c/ t = 2.275\pm 0.125$. This is displayed in Fig. ~\ref{fig: ExtendedAubryAndre} (top panel) as a gray vertical stripe.\\
By performing a similar analysis for different fillings, we obtain the zero-temperature phase diagram as a function of disorder strength and filling factor 
$\rho = N/(2L)$ (see Fig. \ref{fig: ExtendedAAmobilityEdge}). The phase boundary separating the metallic and the insulating phases varies rapidly with the filling. 
Interestingly, these variations are non-monotonic: starting from the zero-density limit, $W_c$ first decreases as the filling increases, then it rapidly increases when the filling is $\rho \gtrsim 0.5$.\\
These findings obtained within the MTIS can be compared with the prediction based on the Anderson criterion of localisation. 
We extract the location of the single-particle mobility edge from the contour plot data of the inverse participation ratio provided 
in Ref.~\citen{Biddle}{}. This procedure is based on the digitalisation of the colour-scale shown in Ref.~\citen{Biddle}, and so it entails some approximations.
Vanishing values of the inverse participation ratio indicate extended single-particle orbitals, while finite values indicate localised states. 
The critical filling factor is obtained when the Fermi energy reaches the mobility edge. 
Notice that in Ref.~\citen{Biddle}{} only the single lattice size $L=500$ was considered, without analysing the finite-size scaling behaviour. 
From the scattering of their data for $L=500$, we estimate the indeterminacy on the extracted critical filling to be close to $10\%$.
Therefore, performing a precise quantitative comparison between our data and those of Ref.~ \citen{Biddle} may not be completely justified. 
However, we see from Fig.~\ref{fig: ExtendedAAmobilityEdge} that the overall agreement is good. In particular, certain important feature of the ground-state phase diagram are 
predicted by both theories. 
First, in the low-filling limit both theories predict the critical disorder to be $W_c/t \simeq 3$, significantly larger than in the native Aubry-Andr\'{e} model (where $W_c/t =  2$). 
Second, the location of the phase boundary has large non-monotonic variations as a function of the filling factor.

%%%%%%%%%%%%%%%%%%%%%%%%%%%%%%%%%%%%%%%%%%%%%%%%%%%%%%%%%%%%%%%%%%%%%%%%
%%%%%%%%%%%%%%%%%%%%%%%%%%%%%%%%%%%%%%%%%%%%%%%%%%%%%%%%%%%%%%%%%%%%%%%%
  \section{1D Anderson model with correlated disorder}
  \label{section5}
        \begin{figure}[ttp!]
 \includegraphics[scale=0.7]{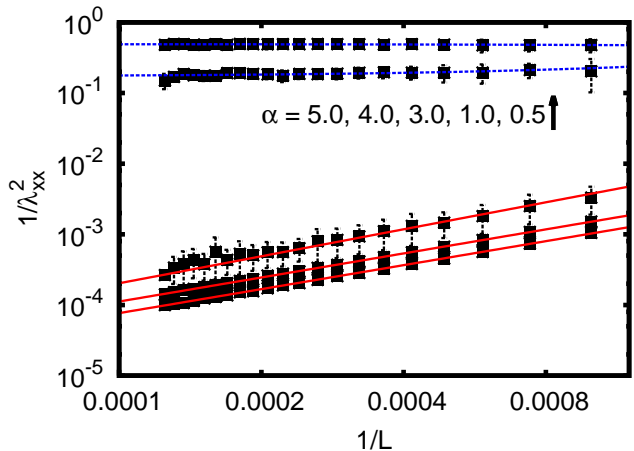}
   \caption{(Colour online) Half-filled 1D Anderson model with long-range correlated disorder 
   and PBC:
   Scaling of ensemble-averaged inverse squared localisation length $1/\lambda_{xx}^2$ as a function of the inverse chain length $1/L$ 
   (log-log scale), for various 
   values of the exponent $\alpha$ characterising the disorder spectral density. 
   For $\alpha = 3.0, 4.0, 5.0$ our simulations indicate a divergence of 
    $\lambda^2_{xx}$ with system size, which is accurately described by the power-law fitting function 
    $\lambda^{-2}_{xx} = cL^{-\gamma}$ ($c$ and $\gamma$ are fitting parameters), 
    shown as  red solid lines.
    For $\alpha = 0.5, 1$ the results suggest a saturation of 
    $\lambda^2_{xx}$ in the thermodynamic limit; the blue dotted lines indicate linear fits.
    }
 \label{fig: CorrelatedDisorder}
\end{figure}
      \begin{figure*}[ttp!]
    \includegraphics[scale=0.7]{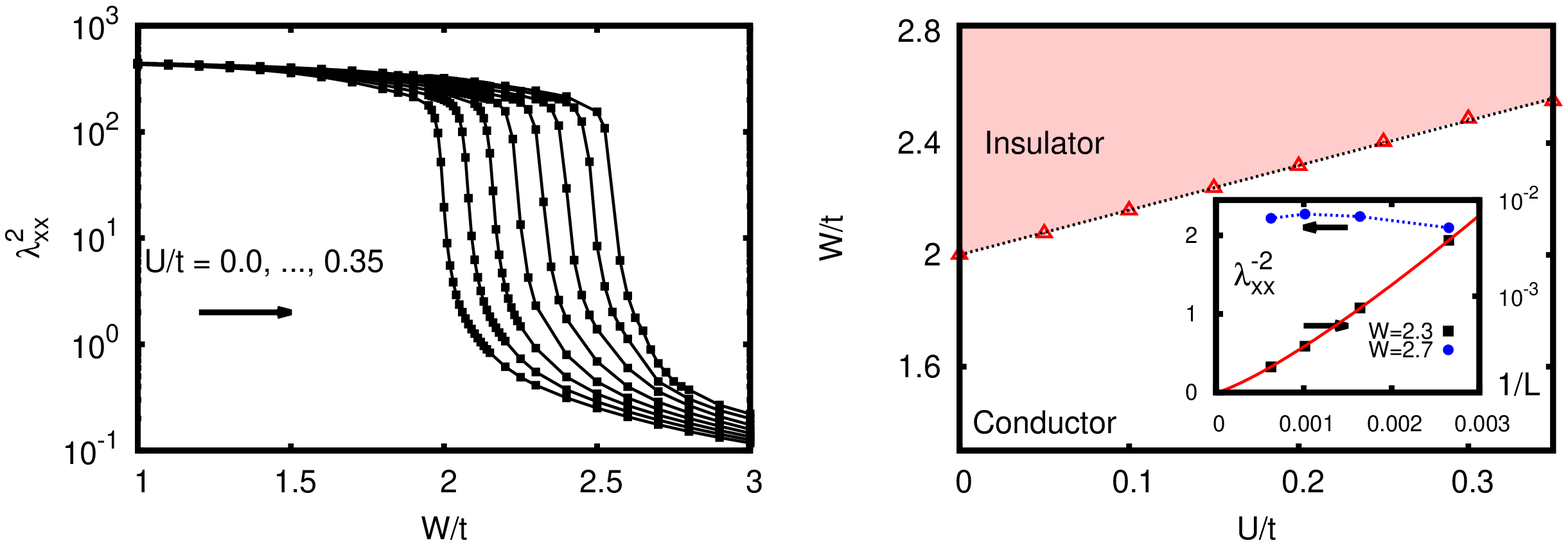}
    \caption{(Colour online) Interacting Aubry-Andr\'{e} model at half filling. 
   Left panel: Squared localisation length $\lambda^2_{xx}$ for a periodic chain of length $L=610$ for various interaction strengths 
   $U$; the jump in $\lambda^2_{xx}$ signals the conductor-insulator transition.
   Right panel: Paramagnetic ground state phase diagram showing the dependence of critical 
   disorder $W_c$ separating insulating and conducting phases as a function of weak interactions $U$. Inset shows scaling of 
   $\lambda^{-2}_{xx}$ with inverse system length $L$ for $U/t=0.3$ at two fixed disorder strengths $W/t = 2.3, 2.7$; the 
   scaling behaviour clearly signals an insulator for the latter and a conductor for the former.
   }
    \label{fig: 1DInteractingAA}
  \end{figure*}
  
  The properties of the Anderson model defined in section~\ref{section2} are affected in a non-trivial manner by the presence of 
  spatial correlations in the disorder pattern, in particular if the correlations have a long-range character~\cite{Soukoulis,Izrlaev,Moura, Croy}.
The effect of long-range spatial correlations has been investigated in several studies, considering in particular disorder patterns characterised by 
  a power-law spectral density $S({k}) \propto k^{-\alpha}$, where $S(k)$ is the Fourier transform of the 
  two-point correlation function $\langle \epsilon_{{r}} \epsilon_{{r}'}\rangle$, and the brackets $\langle \cdots \rangle$ indicate 
  spatial averaging. 
  The value of the exponent $\alpha$ determines the extent of the spatial correlations. The case $\alpha = 0$ corresponds to 
  uncorrelated disorder. 
  For $\alpha > 2$ one has energy sequences with persistent increments~\cite{Moura}.
  A disorder pattern with power-law spectral density can be constructed using the following equation~\cite{Moura}:
  \begin{equation}
   \label{eq: CorrelatedDisorder}
   \epsilon_{{r}} = \sum_{k=1}^{L/2}\left( k^{-{\alpha}}[2\pi/L]^{1-\alpha}\right)^{1/2}\cos{(2\pi {r} k/L + \phi_k)},
  \end{equation}
where $\phi_k$ (with $k=1, \dots L/2$) are random phases sampled from a uniform distribution in the range $[0, 2\pi]$. 
In our calculations, we shift the on-site energies in order to obtain a disorder pattern with zero mean~\cite{Moura}:  
$\epsilon_{\textrm{ave}} = \sum_{{r}} \epsilon_{{r}} = 0$.
Also, in order to curtail the growth of the disorder fluctuations as the system size increases, it is necessary to fix the magnitude of the variance of the 
on-site energies at $ \sum_{{r} } \left(\epsilon_{{r}} - \epsilon_{\textrm{ave}}  \right)^2 = 1$, by appropriately rescaling the on-site energy distribution~\cite{Moura}. 
Notice that, in the notation of Eq.~\eqref{hamiltonian}, the disorder-strength parameter is fixed at $W/t = 1$. This value is not equal to the maximum 
amplitude of the on-site random potential.
This model of correlated disorder is non-deterministic, and so it differs in nature from the deterministic disorder of the  
Aubry-Andr\'{e} models described in Secs. ~\ref{section3} and ~\ref{section4}.
However, its properties are also qualitatively different compared to the uncorrelated Anderson model of section~\ref{section2}.
In fact, a renormalisation group study \cite{Moura} predicted that for large exponents $\alpha > \alpha_c = 2$ the 
single-particle orbitals become extended in a finite portion of the energy spectrum close to the band centre. 
This is in sharp contrast with the case of uncorrelated disorder, where all single-particle orbitals are localised in one-dimensional 
systems~\cite{Gang4,Goldsheidt,Lifshits}.
It is worth emphasizing that the rescaling of the on-site energies was found to be crucial for the occurrence  of the 
single-particle extended states~\cite{Webman,MouraReply,Havlin}.\\
Fig. \ref{fig: CorrelatedDisorder} reports the squared localisation length $\lambda^2_{xx}$ in half-filled chains for various values of 
the exponent $\alpha$. 
The data points correspond to ensemble averages obtained using from $10$  to $200$ realisations of the disorder pattern. The error bars represent the standard deviation of the 
population (instead of the estimated standard deviation of the average) since, as pointed out in Ref.~\citen{Shima}{},  in the presence 
of long-range correlations sample-to-sample fluctuations survive in the thermodynamic limit.\\

As seen from the log-log scale of Fig.~\ref{fig: CorrelatedDisorder}, the datasets corresponding to large exponents $\alpha > \alpha _c$ display a clear 
power-law divergence of the localisation length $\lambda_{xx}$ with the system size, indicating metallic behaviour in the many-particle system.
Similarly to previous sections, we fit the disorder-averaged data for $\alpha  > 2$ with the fitting function 
$\lambda_{xx}^{-2} = c L^{-\gamma}$.
The best fits are obtained with the exponents $\gamma = 1.26(3), 1.130(4), 1.127(4)$, for $\alpha= 3.0, 4.0, 5.0$, 
respectively.
Notice that the latter two data are close to the values found for the Aubry-Andr\'{e} model in the metallic phase. 
In contrast, for lower values of $\alpha<2$ $\lambda$ saturates with the increase of system size. 
This is confirmed by performing linear fits which extrapolate to finite values in the thermodynamic limit.
These findings are consistent with the single-particle analysis of Ref. \citen{Moura}. Indeed, at half filling 
the Fermi energy is close to the band-centre, where the extended single-particle orbitals have been predicted to occur. 
Notice that, according to Ref.~\citen{Shima}, ensemble-averaging causes the transition 
between single-particle localised and delocalised states to morph into a cross-over. However, our aim here is only to identify the 
two many-particle regimes with metallic and insulating phases, without focussing on the precise location of the transition point. 
In fact, we have checked that the two phases (with, respectively, diverging and saturating localisation lengths) can be unambiguously 
identified also by analysing the data corresponding to single realisations (not shown).\\
Our study, from the many-particle perspective of the MTIS, substantiates 
the claim made in Ref.~\citen{Moura} vis-\`a-vis the occurrence of metallic states in 1D.

  %%%%%%%%%%%%%%%%%%%%%%%%%%%%%%%%%%%%%%%%%%%%%%%%%%%%%%%%%%%%%%%%%%%%%%%%
%%%%%%%%%%%%%%%%%%%%%%%%%%%%%%%%%%%%%%%%%%%%%%%%%%%%%%%%%%%%%%%%%%%%%%%%
\section{Interacting Aubry-Andr\'{e} model}
\label{section6}
Understanding the intricate interplay between disorder and interactions in many-fermion systems is an outstanding open problem~\cite{Altshulerbook}. In particular, it is still unclear how an Anderson localized system is affected when electron-electron interactions are included~\cite{Gornyi,Basko}.
The MTIS is clearly a promising approach to address this problem, given that it allows us to describe both noninteracting and interacting insulators within the same formalism.\\
Here we investigate the effect of weak repulsive interactions in the Aubry-Andr\'{e} model. The Hamiltonian we consider is defined as 
\begin{equation}
 \label{eq: 1DInteractingAA}
 H_{\textrm{int.}} = H + U\sum_{r}{n}_{r,\uparrow}{n}_{r,\downarrow},
\end{equation}
where $H$ is the noninteracting Hamiltonian defined in Eq. \eqref{eq: BasicHamiltonian}, with the incommensurate disorder pattern $\epsilon_r$ employed in Secs. \ref{section3}. 
The parameter $U\geqslant 0$ characterizes the interaction strength, while ${n}_{r,\uparrow}$ (${n}_{r,\downarrow}$) is the spin-up (spin-down) density operator at site $r$. 
This Hamiltonian describes the experimental setup recently implemented by the group of Bloch ~\cite{Bloch}. 
The experimenters created quasi one-dimensional tubes with two standing laser waves along the axial direction. 
One of the two lasers has a period which is incommensurate with the other. This creates the deterministic disorder pattern $\epsilon_r$ characterizing the Aubry-Andr\'{e} model. 
The interaction strength $U$ can be tuned employing a Feshbach resonance.\\
Our goal is to determine the zero-temperature phase diagram of the Hamiltonian Eq. \eqref{eq: 1DInteractingAA} at half-filling $N=L$, 
in the regime of weak interactions. We restrict our analysis to paramagnetic phases (allowing charge-density waves), 
and we determine the phase boundary separating the metallic and the insulating ground-states. 
In the regime of relatively strong disorder and weak interactions the Hartree approximation is expected to provide reliable 
results~\cite{Fazekas, Cabra}.
Within this approximation, the Hamiltonian~\eqref{eq: 1DInteractingAA} is simplified using a mean-field decoupling of the interaction 
term, obtaining: 
\begin{eqnarray}
 \label{eq: HFHamiltonian}
 H_{\textrm{int.}} &\approx& H^{\textrm{MF.}, \uparrow} + H^{\textrm{MF.},\downarrow} + I, \nonumber \\
 H^{\textrm{MF.},\uparrow} &=& H^{\uparrow} + U\sum_i \langle n_{r,\downarrow}\rangle n_{r,\uparrow}, \nonumber \\
  H^{\textrm{MF.},\downarrow} &=& H^{\downarrow} + U\sum_i \langle n_{r,\uparrow}\rangle n_{r,\downarrow}, \nonumber \\
I &=& -U\sum_{r}\langle n_{r,\uparrow}\rangle\langle n_{r,\downarrow}\rangle,
  \end{eqnarray}
where $H^{\sigma}$ is that part of $H=H^{\uparrow}+H^{\downarrow}$ corresponding to spin $\sigma=\uparrow,\downarrow$. 
The densities are obtained via a self-consistent iterative procedure based on the equation 
$\langle n_{r,\sigma}\rangle = \sum_{\alpha=1}^{N/2}Q^{(\sigma) *}_{r,\alpha}Q^{(\sigma)}_{r,\alpha}$, where $Q^{(\sigma)}$ is the matrix of 
eigenvectors of the mean-field Hartree Hamiltonians 
$H^{\textrm{MF.}, \sigma}$. Paramagnetism is enforced by setting 
$\langle n_{r,\uparrow}\rangle = \langle n_{r,\downarrow}\rangle$. 
Special care has to be taken in order to ensure that the iterative procedure converges to the true ground-state; following 
Ref.~\citen{Cabra}, we implemented an  annealing scheme where a fictitious temperature parameter is gradually reduced down to zero temperature. 
This temperature-annealing scheme is combined with the standard damping of the density profile provided by each iteration. 
The Hartree formalism is based on an ansatz that the ground state is a Slater determinant, which allows us to 
compute $\lambda^2_{xx}$ as described in Section~\ref{section1}~\cite{RestaTopical}.\\
Our main results are presented in Fig.~\ref{fig: 1DInteractingAA}. The left panel shows the 
squared localisation length $\lambda^2_{xx}$ as a function of disorder strength $W$ for various interaction strengths; the 
sharp jump in $\lambda^2_{xx}$ signals the conductor-insulator transition. 
In the conducting phase the localisation length is cut off by the system size and is independent 
of the interaction strength. 
The inset of the right panel of Fig.~\ref{fig: 1DInteractingAA} shows the scaling of $\lambda^{-2}_{xx}$ with inverse system size for $U/t = 0.3$, 
at two disorder strengths $W/t = 2.3, 2.7$; for the latter value we see that, in the thermodynamic limit, $\lambda^2_{xx}$ 
saturates (signalling an insulator) whereas it diverges for the former (signalling a conductor).\\
An accurate estimate of the transition point is obtained by locating the maximum differential of a polynomial function which fits 
the data of $\lambda^2_{xx}$ as a function of $W$. 
With this procedure we determine the zero-temperature paramagnetic phase diagram of the weakly interacting Aubry-Andr\'{e} model, shown 
in the right panel of Fig.~\ref{fig: 1DInteractingAA}. There is a linear increase of the critical disorder separating the metallic and 
the insulating (paramagnetic) phases. This indicates that repulsive interactions induce delocalisation.
It is worth mentioning that a positive drift of the critical disorder strength has been obtained also in earlier theoretical studies of 
the Mott-Anderson transition in higher dimensions based on dynamical mean-field theory~\cite{Vollhardt}.
An interaction-induced increase of the localisation length was also previously seen in a one-dimensional Anderson-Hubbard model 
within a simple mean-field treatment valid in the atomic limit
~\cite{Henseler}; however, no metallic transition was observed in that study.
In recent experiments a small linear increase of the critical disorder strength was observed in a three-dimensional disordered 
optical lattice~\cite{Marco} . 
Moreover, the bosonic interacting Aubry-Andr\'{e} model was implemented in Ref. \citen{Inguscio}, where interaction-induced delocalisation was
again observed. Bloch's group implemented the fermionic interacting Aubry-Andr\'{e} model~\cite{Bloch}, and determined the 
critical disorder strength where many-body localisation, which is a dynamical phase transition taking place at finite energy density, 
occurs. The critical disorder strength was found to increase as a function of the interaction strength for weak interactions, 
echoing our findings for the ground state. We propose that the 
ground state metal-insulator transition could be observed in their setup by employing the technique used 
in Refs. ~\citen{Marco} and \citen{Tanzi}, where an effective force is imposed on the atoms either by shifting the harmonic 
confinement or by applying a magnetic field gradient; such experimental results could directly be compared with our phase diagram.

%%%%%%%%%%%%%%%%%%%%%%%%%%%
\section{Conclusions}
\label{section7}
Developing approaches to locate insulating transitions in disordered systems, for noninteracting or interacting systems, at zero or finite temperatures, 
is a central problem in condensed matter physics and a subject of intense research~\cite{VollhardtLDOS,Jarrell,Fleishman, Gornyi, Basko, huse}.
In this paper we addressed the zero temperature aspects of this problem from the perspective of the MTIS.
Our findings provide evidence that the many-body localisation tensor, a bulk quantity measuring Kohn's localisation, 
provides a clear signature of the insulating transition induced by disorder at zero temperature.
This was first verified in noninteracting one-dimensional models with uncorrelated disorder, deterministic disorder due to 
incommensurate potentials, and disorder with long-range correlations described by a power-law spectral function.\\
In particular, it was verified that the ground state of the one-dimensional Anderson model is insulating at any disorder strength, in 
agreement with the one-parameter scaling theory~\cite{Gang4} of Anderson localisation.
In the cases of deterministic and correlated disorder, we found evidence of metal-insulator transitions, in agreement with previous 
studies on the critical disorder strength and on the position of the mobility edge based on single-particle theories.\\
Finally we investigated the conductor-insulator transition in a disordered interacting system: 
Using the Hartree mean-field analysis within the MTIS, we found that weak repulsive interactions induce 
delocalisation in the paramagnetic ground state of the Aubry-Andr\'{e} model at half filling, leading to an increase of the 
critical disorder strength required for the onset of insulating behaviour. 
These findings could be observed in cold-atoms using available experimental techniques ~\cite{Bloch,Marco,Tanzi}.\\
One very appealing feature of the present approach is that it permits to identify the insulating
phase using only ground-state properties. On the other hand alternative approaches to identifying the insulating state, such as the Kubo formula for dc conductivity, 
require the knowledge of, at the very least, low-lying excited states. Within the MTIS, the knowledge of the ground-state many-particle wave-function alone suffices, 
a feature which makes it suitable for large scale computational approaches such as, say, quantum Monte Carlo simulations~\cite{Foulkes,Stella}.\\

We acknowledge insightful comments from R. Resta on the manuscript, help from M. Atambo with computing facilities, 
related collaboration with G. Gebreyesus, 
and interesting discussions with T. Nguyen and J. Goold.

\bibliography{Ref}
  
    \end{document}